\documentclass[twocolumn,showpacs,preprintnumbers,amsmath,amssymb,pre]{revtex4}

\usepackage{graphicx}

\begin{document}

\preprint{}

\title{Characteristics of reaction-diffusion on scale-free networks} 

\author{Lazaros K. Gallos}
\author{Panos Argyrakis}
\affiliation{Department of Physics, University of Thessaloniki, 54124 Thessaloniki, Greece}

\date{\today}

\begin{abstract}
In this work we examine some characteristic properties of reaction-diffusion processes of the $A+A\to 0$ type on scale-free networks.
Due to the inhomogeneity of the structure of the substrate, as compared to usual lattices,
we focus on the characteristics of the nodes where the annihilations occur.
We show that at early times the majority of these events take place on low-connectivity nodes,
while as time advances the process moves towards the high-connectivity nodes, the so-called hubs.
This pattern remarkably accelerates the annihilation of the particles, and it is in agreement
with earlier predictions that the rates of reaction-diffusion processes on scale-free networks
are much faster than the equivalent ones on lattice systems.
\end{abstract}


\maketitle

\section{Introduction}

The $A+A\to 0$ reaction is one of the most fundamental annihilation processes, with many applications
\cite{benBook}.
In a number of papers \cite{GA04,CBPS05a,GA05} this reaction was studied on scale-free networks,
both computationally and analytically. 
In a recent Letter \cite{GA04} we reported some unusual properties of reaction-diffusion
processes taking place on scale-free networks. These networks represent a special class of
random networks, where the probability that a node is connected to $k$ other nodes of the
network follows a power-law form:
\begin{equation}
\label{Pk}
P(k) \sim k^{-\gamma} \;,
\end{equation}
where the value of the exponent $\gamma$ usually lies in the range $2<\gamma<4$. A large number of systems from
markedly different disciplines have been found to fall in this class, and new systems
are added continuously in the list \cite{AB02,DM02}. This ubiquity explains the intense interest devoted
to the study of the complex networks field. The unusual structure of the scale-free networks
has been shown to strongly modify the properties of dynamic processes, where particles move from node
to node along the existing links, as compared to the same heavily-studied processes on lattice
systems or continuous space.

Reaction-diffusion mechanisms are a frequently encountered mechanism for kinetics, where particles
diffuse in a space and react in a predefined way upon encountering other particles.
For example, during the $A+A\to 0$ process, particles of the $A$ type diffuse and
annihilate when they collide with other particles of the same type.
In Euclidean lattices, the simplest mean-field approach predicts a linear increase
of the inverse particle concentration $\rho(t)$ with time $t$. A large number of studies \cite{ZO78,TM83,TW83},
though, demonstrated that the density actually scales as
\begin{equation}
\frac{1}{\rho(t)} - \frac{1}{\rho_0} \sim t^f \,,
\end{equation}
where $f<1$, and $\rho_0$ is the initial particle concentration at time $t=0$. In a space
with dimensionality $d$ this exponent equals $f=d/d_c$, for $d\leq d_c$ and $f=1$ for
$d> d_c$. The critical dimension for the $A+A$ reaction is $d_c=2$.
This `anomalous' behavior has also been observed when the substrate of diffusion has
different geometry, such as a fractal structure \cite{bAH87}, where $f=d_s/d_c$, and $d_s$ is the spectral
dimension.

The maximum value of $f$ in all the above cases has been $f=1$. In scale-free networks, though,
we recently found numerically that the process follows a different mechanism. For networks with $\gamma\leq 3.5$, the concentration decay still follows a
power law, which for a short interval is described initially by $f<1$, but soon after that
exhibits a crossover to a power law with an exponent $f>1$. This is a clear indication for
the rapid acceleration of the process, which we attributed mainly to the existence of the hubs.
Following this, Catanzaro et al. \cite{CBPS05a} developed a theoretical framework for `uncorrelated' networks
(i.e. networks with no degree-degree correlations, see \cite{CBPS05b}). They predict a behavior
\begin{equation}
\frac{1}{\rho(t)} \sim \left\{
\begin{array}{cc}
t^{1/(\gamma-2)}, & 2<\gamma<3 \\
\ln(t), &  \gamma=3 \\
t ,& \gamma>3
\end{array}
\right. \,.
\end{equation}
A detailed comparison between the two models was recently presented in \cite{GA05}, where it also
became evident that the exact behavior of the particle concentration depends on
details of the structure, such as, for example, the minimum number of links on a node.

In the present work we study the reaction-diffusion $A+A\to 0$ model on
scale-free networks, with emphasis on the nodes where the annihilation events take place on,
and we try to better understand the mechanism that leads to this new type of behavior.

\section{The model}

The networks that we use in our simulations are prepared with the standard configuration model \cite{MR95}.
Initially, an integer $k$ value representing the number of links
is assigned from a random distribution obeying Eq.~\ref{Pk}
to every one of the $N$ nodes in the network. Random pairs of links are chosen and connect
two nodes, but double links and self-connections are not allowed. This network creation method
may create a certain number of separated clusters (depending on the value of $\gamma$). We isolate and
use only the largest cluster of the network, where the particles are allowed to diffuse.

During the reaction-diffusion process
the number of particles $M(t)$ in the system is reduced with time, and we denote with $\rho(t)$ the
particle concentration at time $t$.
The initial particle concentration is $\rho_0$. A number of $N_0 = \rho_0 N$ particles are placed
on randomly selected nodes (in simulations of this work we use $\rho_0=0.5$). Then, we randomly choose one particle and pick one of the neighboring
nodes where this particle is located at. If this new node is empty, the particle moves and occupies its new position.
If this node is already occupied, then the two particles annihilate and are immediately removed from
the system.
Time is advanced inversely proportionally to the current number of particles, by $1/M(t)$. We repeat this procedure
by continuously selecting, moving and (possibly) annihilating particles. We perform
100 different realizations of the walk on different networks of size $10^6$ nodes each.
The largest cluster on these networks is, of course, smaller and spans 35\% to 100\% of the
system nodes, depending on the value of $\gamma$.

\section{Results}

In order to gain insight into the reaction process, and given that the substrate of the diffusion is
very inhomogeneous, we try to understand the nature of the nodes where the annihilations take place. The connectivity 
of the nodes in the network varies largely from node to node, but the great majority has a small
number of connections, which is the main characteristic of a scale-free network. A very small number of
nodes are well-connected (the hubs), and the existence of the hubs
has been shown to be the main reason behind most of the unusual properties reported for the scale-free
networks.

\begin{figure}
\begin{center}
\includegraphics{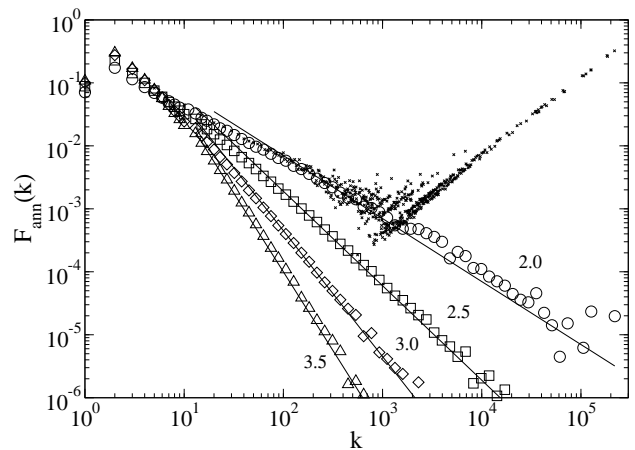}
\end{center}
\caption{\label{fig1} Probability, $F_{\rm ann}(k)$, for an annihilation event to take place on a node with $k$ links
as a function of $k$, during a process of $t=100$ steps. Open symbols are averages over 100 scale-free networks
with $\gamma$=2.0, 2.5, 3.0 and 3.5 (top to bottom, shown in the plot), of $N=10^6$ nodes each. The data have been
logarithmically binned for $k>10$. Lines represent Eq.~\protect\ref{EQmf},
with the corresponding exponent $1-\gamma$ for each case. We also present unbinned data (dots) for the case of $\gamma=2.0$.}
\end{figure}

In this context, we have performed simulations of the $A+A$ reaction-diffusion problem on scale-free
networks of various $\gamma$ exponents. After $t=100$ steps, when the particle concentration has diminished
to small values, we calculate the percentage of annihilation events $F_{\rm ann}(k)$ that took place
on a node with $k$ links. Figure \ref{fig1} shows $F_{\rm ann}(k)$ as a function of $k$.
The raw data for $F_{\rm ann}$ exhibit a kink at values of $k$ around 100 to 1000, above which
the curve increases monotonically with increasing $k$. We present one such curve in Fig.~\ref{fig1}
for the case of $\gamma=2.0$. This feature (which is also size-dependent) is
a numerical artifact, due to the rarity of nodes with very large degree $k$, and is removed when the data are
binned.
The four curves in the figure are the result of logarithmic binning at degrees $k>10$.

In a mean-field treatment of a network with no degree-degree correlation we expect that the number of events would be proportional
to the total number of links leading to a $k$-node, i.e.
\begin{equation}
\label{EQmf}
F_{\rm ann}(k) \sim k P(k) \sim k^{1-\gamma} \,.
\end{equation}
The exponent $1-\gamma$ is verified in all four cases, as can be seen in Fig. \ref{fig1}, for the intermediate to large $k$ regime.
The result of Eq.~\ref{EQmf} and the presented curves show that even though most of the events take place on low-$k$ nodes,
which comprise most of the network, the probability for annihilation at the hubs is significantly larger than their relative appearance in a network.
The behavior seems also not to be influenced by any degree-degree correlation in the network. The probability that nodes with low connectivity are connected to
a hub can in fact be larger than the one predicted for a completely uniform distribution \cite{CBPS05b}, due to the avoidance of double links
and self-connectivity.

\begin{figure}
\begin{center}
\includegraphics{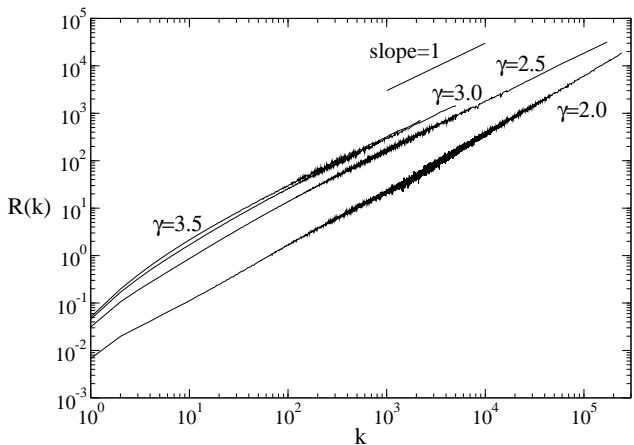}
\end{center}
\caption{\label{fig2} Average number of annihilation events per node, $R(k)$, with $k$ links as a function of $k$,
during a process of $t=100$ steps. Curves are averages over 100 scale-free networks
with $\gamma$=2.0, 2.5, 3.0 and 3.5, of $N=10^6$ nodes each.}
\end{figure}

The importance of the hubs can be made explicit if we ask what is the average number of events $R(k)$ on a single node with
$k$ links, i.e.
\begin{equation}
R(k) = \frac{F_{\rm ann}(k)}{P(k)} \,.
\end{equation}
According to Eq. \ref{EQmf} we expect a linear increase with the number of links $k$, since $R(k)\sim k^{1-\gamma}/k^{-\gamma}
\sim k$. This behavior is verified in Fig.~\ref{fig2}, at least for large $k$ values. The slope of $\gamma=2$ seems
to be slightly different than the other curves, but this can be attributed to the increasing number of hubs at this $\gamma$ value,
where the network has a very small diameter. The curves in Fig.~\ref{fig2} span many decades on the $y$-axis manifesting a very different
role of nodes with different degrees. The case of $\gamma=2$ is also the borderline value
that separates networks with finite degree distribution moments from networks with
infinite moments. The diameter of networks with $\gamma<2$ is extremely small and the
hubs connect practically the entire network within a few steps. In such networks with
$\gamma<2$, which are difficult to simulate numerically, we expect that the annihilation process will be exponentially fast and that almost all the reaction activity will take place
on the hubs or on their immediate neighbors.

In a typical walk of $t=100$ steps about $2\times 10^5$ events are observed. If we focus on
a particular node with a small degree we see that there is much less than one annihilation event occuring on this node during the process. On the contrary,
more than $10^4$ events occur at the hubs having $k\simeq 10^5$ links. We, thus, conclude from Figures \ref{fig1} and \ref{fig2}
that a hub is a much more active element in a network than any other node, although the majority of events still occurs over the
large number of low-$k$ nodes.

\begin{figure}
\begin{center}
\includegraphics{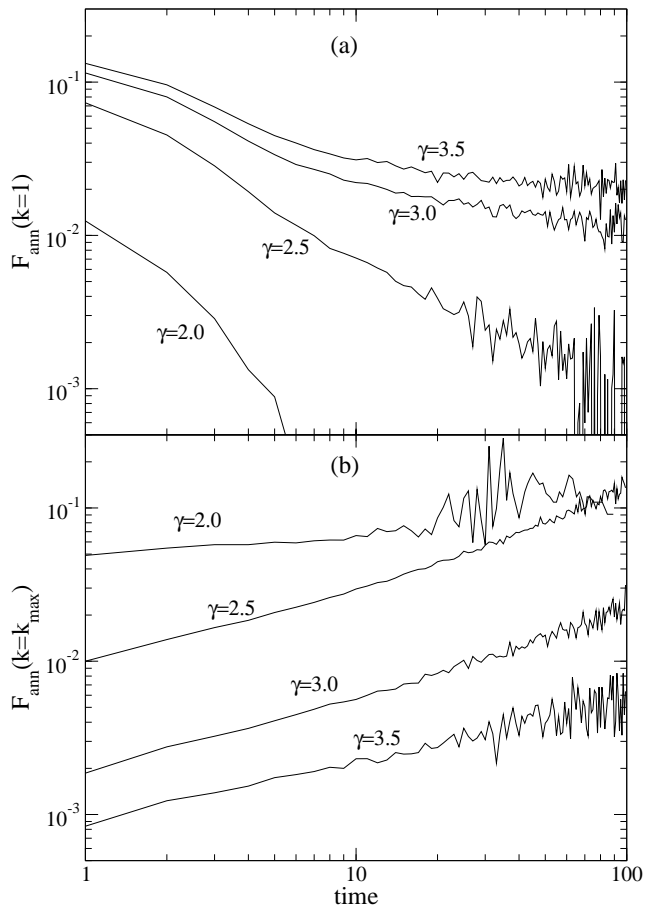}
\end{center}
\caption{\label{fig3} Time evolution of the average probability (a) $F_{\rm ann}(k=1)$ and (b) $F_{\rm ann}(k=k_{\rm max})$,
for networks with varying $\gamma$ exponents.}
\end{figure}

The results until now concerned the entire process after averaging over the first $t=100$ steps. We now turn to the time
evolution of the diffusion-reaction mechanism. In Fig.~\ref{fig3}a we can see how $F_{\rm ann}(k=1)$ varies with time.
For all $\gamma$ values the probability starts initially with a certain value and decreases monotonically with time before
stabilizing to a much lower value. For $\gamma=3$ and $\gamma=3.5$, where hubs are not very strongly connected, the probability
goes down to a value of $0.02$ even for the longer times displayed. For $\gamma=2$ and $\gamma=2.5$, though, the probability
that an annihilation occurs on a node with $k=1$ practically vanishes. This picture is reversed when we monitor the 
probability $F_{\rm ann}(k_{\rm max})$, i.e. the fraction of events taking place on the single most connected node of the network
during the time interval $[t,t+1]$ over the total number of events that occured during the same interval. Figure~\ref{fig3}b
suggests that now $F_{\rm ann}(k_{\rm max})$ starts with low values but continuously increases as time evolves.
Note that for networks with $\gamma<3$ more than 10\% of the events at long times take place on this node alone.

\begin{figure}
\begin{center}
\includegraphics{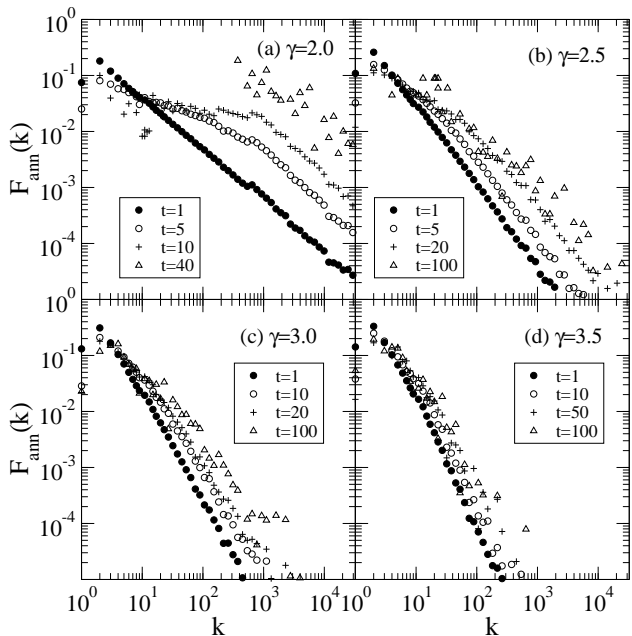}
\end{center}
\caption{\label{fig4} Time evolution of the probability distribution $F_{\rm ann}(k)$ for four different $\gamma$ values (shown on the plot).
Data have been logarithmically binned for $k>10$.
Each curve corresponds to different times displayed in the legends.}
\end{figure}

In Fig.~\ref{fig4} we present the time evolution of the entire $F_{\rm ann}(k)$ distribution for different
$\gamma$ values. The distribution at $t=1$ has a form similar to that of Fig. \ref{fig1}.
When $\gamma=2$, and as time
evolves, the low-$k$ part decreases and the large-$k$ part increases. Asymptotically, the distribution tends to uniform and
for $t\simeq 40$ there is no event taking place at nodes of small connectivity. For $\gamma=2.5$ the
picture is similar, but at $t=100$ the distribution has not yet spread uniformly over the entire $k$ range. For $\gamma\geq 3$
the right wing increases with a slower rate than when $\gamma<3$, and always remains significantly lower than the small-$k$ part,
at least for the early time scale that we observe in our current work.

Notice also that at time $t=100$ the process on networks with different $\gamma$ lies in different stages of the
$\rho(t)$ evolution, as can be seen in Fig.~1 of Ref.~\cite{GA04}. For $\gamma=2$ the concentration has almost
diminished to zero, while for $\gamma=2.5$ we are well inside the power law regime with $f>1$. When $\gamma=3.0$
we are located close to the crossover point between the two power law regimes, and for $\gamma=3.5$ the concentration
has not reached the transition point and still decays slower than linear.

In Ref.~\cite{GA04} we had shown that during an $A+A\to 0$ reaction particles tend to segregate. This is a very unusual
type of behavior, since in regular Euclidean space the opposite effect (formation of depletion zone) is observed.
We, thus, consider in more detail the cluster structure of the particles in the network. We have already
seen (Fig.~3a of \cite{GA04}) that the fraction $Q_{AA}$, defined as the number of the close contacts in the system over the total possible
number of contacts, increases as a function of time (except for $\gamma>3$), a clear indication towards particle
segregation.
In order to fully characterize the motion and the spatial distribution of the particles we now use the two-particle correlation function $g(L)$,
which we define as the number of particles at distance $L$ over the number of particles at distance $L$ when we assume
an equal but uniform particle distribution. On random networks, this random distribution cannot be accurately estimated
theoretically, as e.g. for simple lattices, due to the widely different local environment. In order to overcome
this problem, for every network that we used and for every concentration encountered we placed a corresponding
number of particles uniformly on this network and averaged over 20 different realizations. By calculating the
average number of particles at distances $L$ we were able to determine the value of the denominator.
Thus, if the particles are randomly distributed in the network this quantity will be 1.
If the particles tend to occupy nodes far from each other this fraction will be $g(L)<1$, while $g(L)>1$ shows that particles
are clustered in close contact to each other.

\begin{figure}
\begin{center}
\includegraphics{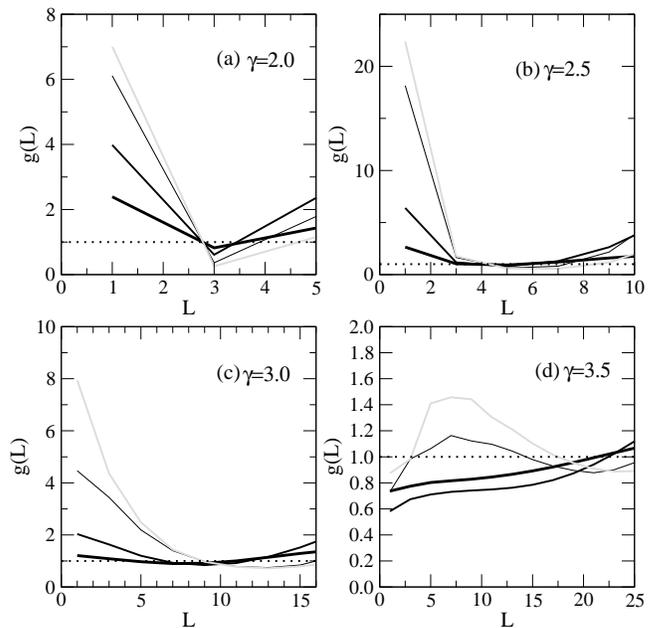}
\end{center}
\caption{\label{fig5} Two-particle correlation function $g(L)$ as a function of the distance $L$ for different $\gamma$ values (shown in the plot).
The dashed line represents the uniform distribution limit $g(L)=1$.
Thicker curves correspond to earlier times. Times used in each plot: (a) $t=$3, 8, 20, 30. (b) $t=$6, 25, 100, 150,
(c) $t=$12, 72, 500, 850, and (d) $t=$15, 140, 1800, 3400.}
\end{figure}

The first three plots in Fig.~\ref{fig5} (corresponding to $\gamma\leq 3$) describe a picture where particles break
their initial uniform distribution and cluster with an increasing rate. The average number of neighbors within a few
steps is significantly high and increases with time, while the number of particles at longer distances are close to
the mean-field assumption $g(L)=1$. When $\gamma=3.5$ the correlation function is more similar to a typical $A+A$ reaction
on a normal lattice.
Initially, a depletion zone is created around the A particles. This depletion zone is retained for the duration of the
process, with the difference that at longer times we observe an increase of particles concentration at moderate distances.
This increase is, though, much smaller than for the case of $\gamma\leq 3$.

\section{Summary}

Summarizing, in the present study we focused on the details of the diffusion,
such as the spatial organization of the particles and the time evolution leading to an annihilation
reaction, with a scale-free network as the substrate.

The majority of the particles annihilate on the low connectivity nodes. This happens mainly at
early times, where the static distribution of particles on the network is the dominant factor.
The particles annihilate in the vicinity of their initial neighborhood, which on average includes mostly
low-degree nodes (since the number of such nodes in the system is large).

At later times, when a small number of particles remain in the system, diffusion starts to play an
increasingly important role. Particles encounter each other mainly on nodes where diffusion directs them to,
i.e. on the hubs\cite{Gallos}. Thus, although the percentage of the hubs in the network is very small
most reactions take place on them and they dominate the process at longer times.

\end{document}